\documentclass[preprint,12pt]{elsarticle}

\usepackage{amssymb}

\usepackage{placeins}

\usepackage{mathrsfs}

\usepackage{subfig}

\journal{International Journal of Multiphase Flow}

\begin{document}

\begin{frontmatter}



\title{An Analysis of the Convergence of Stochastic Lagrangian/Eulerian 
Spray Simulations}


\author[label1]{David P. Schmidt\corref{cor1}} \ead{schmidt@acad.umass.edu}

\author[label2]{Frederick Bedford} \ead{fritz.bedford@siemens.com}
\address[label1]{Mechanical and Industrial Engineering Department \\
                 University of Massachusetts Amherst, MA}
\address[label2]{Siemens Product Lifecycle Management Software Inc., Lebanon, NH}

\begin{abstract}
This work derives how the convergence of stochastic Lagrangian/Eulerian 
simulations depends on the number of computational parcels, particularly for 
the case of spray modeling.  A new, simple, formula is derived that can be used 
for managing the numerical error in two or three dimensional 
computational studies.  For example, keeping the number of parcels per cell 
constant as the mesh is refined yields an order one-half convergence rate in 
transient spray simulations.  First order convergence would require a 
doubling of the number of parcels per cell when the cell size is halved.  Second 
order convergence would require increasing the number of parcels per cell by a 
factor of eight.  The results show that controlling statistical error 
requires dramatically larger numbers of parcels than have typically been used, 
which explains why convergence has been so elusive.  
\end{abstract}


\end{frontmatter}


\section{Introduction to the Problem}

The computational fluid dynamics community expects numerical schemes to be 
consistent, i.e. they should converge to the exact solution as resolution 
is increased.  Without this assurance, the modeler can wander aimlessly, 
confused by pathological numerical errors.  The present work will identify 
a simple rule that governs convergence and will bring Lagrangian/Eulerian spray 
simulation closer to modern standards of simulation quality. The mathematical 
analysis is intended to also benefit other forms of Lagrangian/Eulerian 
simulations, such as bubbly or particle-laden flow.

The first step is to identify what is and what is not understood.  The typical 
parts of a Lagrangian/Eulerian spray computation can be categorized as follows:

\begin{enumerate}
\item {Gas phase equations:} The gas phase conservation equations, excluding 
source terms for interaction with the spray, solved in an Eulerian frame of reference
\item {Particle tracking:}  The governing ordinary differential equations that 
describe the evolution of the tracked particles, solved in a Lagrangian frame of reference
\item {Gas to liquid coupling:}  The interpolation of gas phase quantities to 
the location of the spray parcels
\item {Liquid to gas coupling:} The agglomeration of the scattered point 
particle effects into source terms in the continuous phase equations
\end{enumerate}

The first item is well understood.  The second category includes the models 
that describe the evolution of droplets as they move in the Lagrangian reference 
frame.  Are et al.\ \cite{Are} provided analysis for the basic equations of 
spray motion.  Spray breakup has not been analyzed but is expected, for at least 
basic breakup models, to behave as ordinary differential equations (ODEs) which 
are well understood.  Droplet collision is beyond the scope of this paper 
and has been studied elsewhere \cite{Schmidt2000, Abani2008}.  Gas to liquid 
coupling is equivalent to interpolation and was also analyzed by Are et al.

The final item, representing the effect of the liquid phase on the gas phase, 
has proven to be the most challenging.  Here, the numerical methods accumulate 
information from the liquid phase about mass, momentum, energy, and species 
contributions to the gas phase.  These methods must contend with the fact that 
parcels are not collocated with gas phase nodes or finite volume cell centers.  
Further, because of the limited number of parcels used in spray computations, 
there is a statistical uncertainty in these terms.

\begin{figure}
 \centering
 \includegraphics[width=0.85\textwidth]{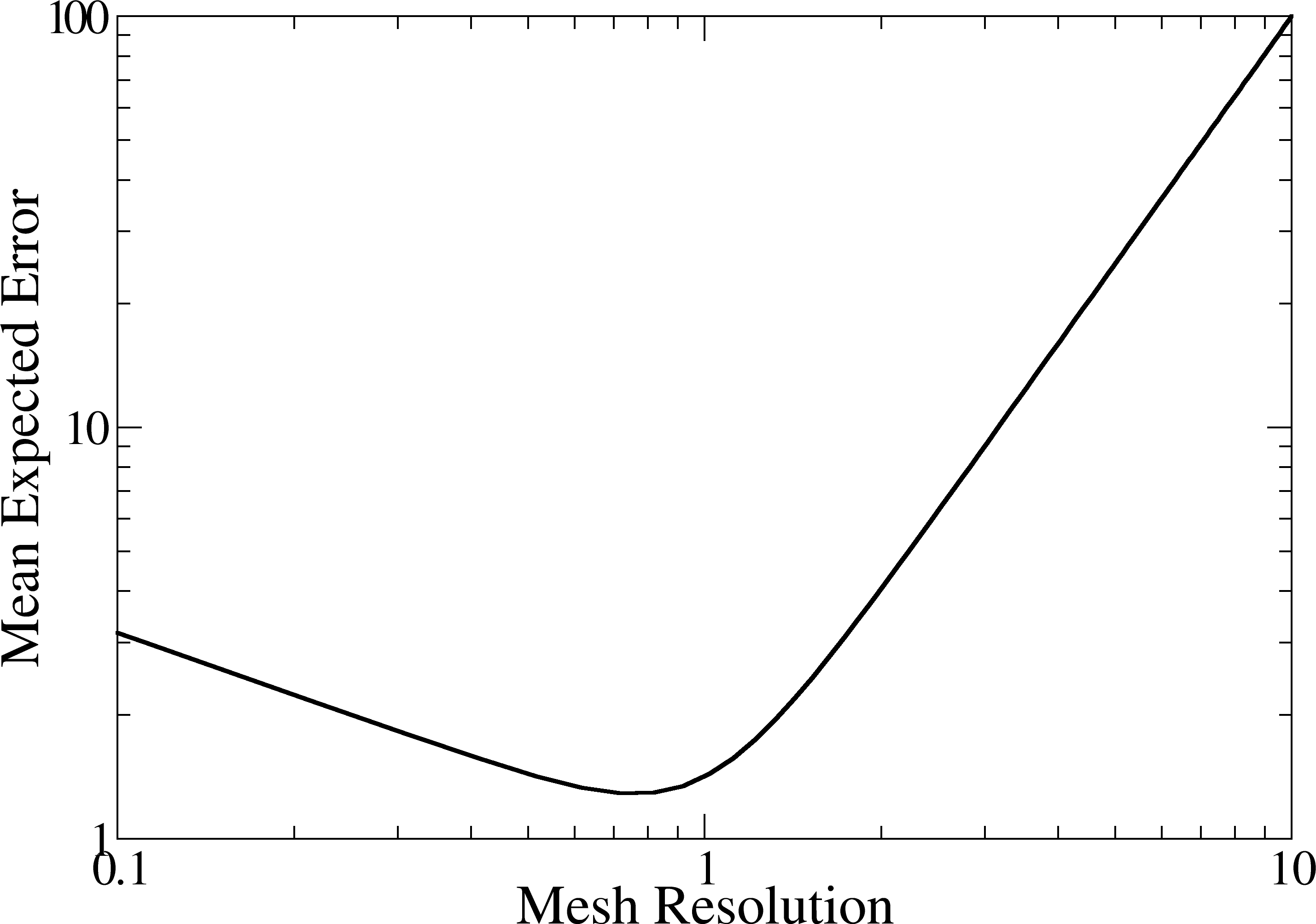}
 \caption{General behavior of error when the number of parcels is held constant. 
This situation corresponds to employing Eqn.~\ref{eq:L2} with a value of $a=0$.}
 \label{fig:optimum}
\end{figure}

Because of this statistical uncertainty, refinement of gas phase solution without 
adequately refining the Lagrangian solution by increasing the number of computational 
parcels can actually result in a less accurate answer.  Early work by Subramaniam 
and O'Rourke \cite{Subramaniam} concluded that the convergence of spray simulations 
was conditional on employing a sufficient number of parcels.  Subsequent analysis 
by Schmidt \cite{Schmidt2006} saw similar trends.  If the number of parcels is held 
constant while the mesh is refined, the error begins to increase after passing an 
optimum, as shown in Fig.~\ref{fig:optimum}.

There have been several past investigations of liquid to gas coupling.  The 
work of Stalsberg-Zarling et al.~\cite{stalsberg2004} applied Lagrange 
polynomial interpolation to interphase coupling in order to produce a more 
stable, less mesh-dependent simulation.  The work of Garg et al.\ \cite{Garg07} 
used several problems that admitted analytical solutions in order to empirically 
study the convergence rate of fairly sophisticated coupling schemes. For completeness, 
they included statistical factors in their tests by varying the number of parcels per 
cell.  In order to understand the behavior and causes of the error, they modeled 
deterministic, bias, and statistical contributions using a framework suggested by 
Dreeben and Pope \cite{Pope}.  

The work of Garg et al.\ judged the efficacy of these liquid to gas coupling schemes 
based on empirical tests. In later work, Garg et al.\ \cite{Garg09} developed a 
parcel number density control system to moderate statistical error and, with the 
help of a sophisticated liquid-to-gas source estimation scheme, produced 
convergence of a fully-coupled Lagrangian/Eulerian calculation.  Their success 
strongly supports the notion that the distribution of sources from the Lagrangian 
phase to the Eulerian is the crux of controlling overall error.
 
In contrast, the work of Are et al.\ excluded considerations of statistical 
error in both their analytical and empirical tests.  The analysis took the 
limit of an infinite number of parcels, while the empirical tests used 
arbitrary, large numbers of parcels.  They succeeded in showing theoretically, for 
each of the elements listed above, how to achieve second order accuracy.  They then 
combined these elements into a full Lagrangian/Eulerian simulation and demonstrated 
second order convergence. Later, Schmidt \cite{Schmidt2006} studied how second order 
and higher accuracy schemes converge considering both spatial and statistical 
error.  However, Schmidt only considered liquid-to-gas coupling and only for fixed 
numbers of parcels per cell.  

The outstanding question that has never been answered is:\textit{``In a 
Lagrangian/Eulerian simulation, how fast must the number of parcels 
increase in order to observe convergence during mesh refinement?''}  Without 
knowing the answer, one cannot reliably demonstrate spray calculation 
convergence.  The present work derives a recipe that answers this question 
and performs demonstrations in order to empirically confirm the predictions.    

This effort will add to the body of literature that has empirically studied 
convergence, such as Senecal et al.\ \cite{Senecal2012}.  These studies have 
proceeded without analysis to provide guidance.  Senecal et al.\ is an 
example where the authors successfully achieved convergence by a combination of 
careful numerical methods and employing a substantial number of computational 
parcels.  In a subsequent section of this paper, their approach will be 
compared to the rule which we derive.  

\section{Analysis of a Single Time Step}

Our analysis examines the calculation of particle to gas source terms as a
function of location in a $d$ dimensional space.  The magnitude of 
the sources in this analysis is determined by the gas-to-particle coupling, 
reducing the particle-to-gas coupling to a determination of how the sources 
are to be distributed on the gas phase mesh.  Consequentially, the distribution 
of sources can be represented by a density function
$f \left( \vec{x} \right) $ which is estimated from $n$ non-uniformly spaced parcels 
(See Pai and Subramaniam \cite{Pai} for an investigation of 
discrete particle representations of smooth distribution functions).  
For each continuous phase cell, the contributions of the local parcel sample 
is summed using a kernel with compact support. 

Our approach will be to construct an estimate of the combined error due to 
statistical and spatial contributions.  This estimate will be dependent on both 
the number of parcels and the level of spatial resolution.  We will then assert 
that the number of parcels must be tied to the cell size by a simple power law 
relation, which then results in an error estimate that is only a function of 
cell size.  Finally, we will calculate the convergence rate of this error.

The analysis begins with our representation of the kernel used to connect 
parcel contributions to the source term.  As reviewed by Garg et al.\ 
\cite{Garg07} and Schmidt \cite{Schmidt2006} there are several approaches that 
can achieve varying degrees of spatial accuracy or reduced statistical error.
The simplest and most common approach in Lagrangian/Eulerian spray simulations 
is the nearest-node kernel.  This treatment, in a finite volume context, is 
simply:  if a parcel resides in a cell, then all of the contributions from the 
parcel go the gas phase equation for that cell. 

The nearest node kernel can be represented as a function $K$ that depends on 
the dimensionless distance $y$ from the center of the cell in a 
$d$-dimensional Cartesian space, where $m$ is the index of a dimension.

\begin{equation}
 \label{eq:kernel}
 K_m \left( y \right) = 1
\,\,\,\,\,\,\,\,\,\,\,\,\,\,\,\,\,\,\,\,\, 
\left| y \right| \leq \frac{1}{2}
\end{equation}

The present work will consider the simplified case where the kernel is the same 
in each dimension and the mesh spacing $\Delta x_m$ is uniform and constant in 
each dimension, and so the subscript $m$ is dropped.  The argument of the kernel, 
$y$, is defined for the $j_{th}$ parcel below, where $\vec{x}$ is the location 
of the nearest cell center.  Similarly, $\vec{x}_j$ is the 
location of the parcel.  The symbol $\hat{m}$ represents a unit vector in the 
$m$ direction.

\begin{equation}
 y=\frac{ \left( \vec{x} -\vec{x}_j \right) \cdot \hat{m} }{\Delta x}
\end{equation}

The numerical estimate of the source term distribution function, 
$f_n \left( \vec{x} \right) $, is the result of applying the kernel in all 
dimensions to all parcels.  Here, $n$ represents the number of parcels in the 
entire domain.

\begin{equation}
\label{eq:fn}
 f_n \left( \vec{x} \right) = \frac{1}{n \Delta x^d} \sum_{j=1}^{n} {
\prod_{m=1}^{d} {K_m \left( \frac{ \left( \vec{x} -\vec{x}_j \right) \cdot 
\hat{m} }{\Delta x}  \right) }}
\end{equation}

The mean square error, $E^2$, is defined as the square of the difference 
between the true source distribution function, $f$, and the expected numerical 
value, $f_n$, produced by Eqn.~\ref{eq:fn}.  Here, we can build on prior efforts 
to understand the related problem of finding the optimal kernel for estimating density 
functions.  Epanechnikov \cite{Epanechnikov} calculated the mean square error for 
a $d$ dimensional space and a general kernel.

\begin{equation}
 \label{eq:mse}
E^2 = \frac{f}{n} \left[ \prod_{m=1}^{d} \frac{1}{\Delta x} 
\int_{-\infty}^{\infty} K^2_m \left( y \right) dy \right]
+
\frac{1}{4} \left[ \sum_{m=1}^{d} \frac{ \partial^2 f}{\partial x_m^2} \Delta 
x^2 \right]^2
\end{equation}

The first term represents the statistical error, which depends on the variance 
of the kernel and can be calculated in each dimension separately for the 
nearest node kernel \cite{Rosenblatt}.

\begin{equation}
 \int_{-\infty}^{\infty} K^2_m \left( y \right) dy = 1
\end{equation}

Inserting the nearest-node kernel definition in the mean square error estimate 
gives the specific result in Eqn.~\ref{eq:mse_nn}.  It is comprised of two terms, 
the first representing the stochastic error in the source term and the second 
representing the spatial error.  In the case considered here, for very small 
$\Delta x$ the first term will dominate unless $n$ increases sufficiently fast 
with decreasing cell size.  For situations where the cell size is large compared 
to the length scale established by the second derivative, the spatial error term 
may instead dominate.  

\begin{equation}
 \label{eq:mse_nn}
E^2 = \frac{f}{n \left( \Delta x \right)^d} 
+
\frac{1}{4} \left[ \sum_{m=1}^{d} \frac{ \partial^2 f}{\partial x_m^2} \Delta 
x^2 \right]^2
\end{equation}

The mean square error is a point-wise measurement that can be assessed over the 
whole domain using the $L_2$ norm. For a discrete simulation, $L_2$ is calculated 
as follows.

\begin{equation}
 \label{eq:L2def}
 L_2^2 = {\Delta x}^d \sum_{cells} E^2
\end{equation}

Now the goal is to find how the number of parcels in the domain, $n$, must 
depend on $\Delta x$ such that error measured by $L_2$ decreases at a desired 
rate.  The relationship between $n$ and the cell size $\Delta x$ is 
stipulated as a power law relationship as given by Eqn.~\ref{eq:model}
\begin{equation}
 \label{eq:model}
 n = \frac{b}{ \Delta x ^{a} }
\end{equation}
where $b$ is an arbitrary constant.  The parameter $a$ controls the rate at 
which the number of parcels must increase with mesh refinement.  For example, 
the case of $a=0$ indicates no increase in the numbers of parcels with mesh 
refinement.  Setting $a=d$ means that the number of parcels per cell remains 
constant with mesh refinement.

Combining the last three equations gives the following result.

 \begin{equation} 
 \label{eq:L2sum}
 L_2 \approx
 \sqrt{
 \frac{\Delta x^{a-d} }{b} \sum_{cells} f \Delta x^d 
 + 
 \frac{\Delta x^4}{4}  \sum_{cells}
 \left[
 \sum_{m=1}^{d} \frac{ \partial^2 
f}{\partial x_m^2}
\right]^2
\Delta x^d
 }
 \end{equation}
where the integer $m$ indexes the dimensions up to $d$ in the second term.
Allowing $\Delta x$ to become arbitrarily small gives the limiting behavior:

 \begin{equation}
 \label{eq:L2}
 L_2 \approx 
 \sqrt{
 \frac{\Delta x^{a-d} }{b} \int_{x \in R^d} f dx^d 
 + 
 \frac{\Delta x^4}{4}  \int_{x \in R^d}  
 \left[
 \sum_{m=1}^{d} \frac{ \partial^2 
f}{\partial x_m^2}
\right]^2
dx^d
 }
 \end{equation}

The first term under the radical represents statistical error and the second 
term represents spatial error.  These two sources are in contrast to the three 
error contributions suggested by Dreeben and Pope \cite{Pope}.  The present 
analysis results in an unbiased estimation of the source terms.  

However, this analysis should not be interpreted as a dismissal of bias error.  
The present analysis does not include the full set of evolution equations, so the 
potential still remains for bias to appear through feedback and propagation in the 
droplet motion calculation and the Navier-Stokes equations.
 
The quantities inside the two integrals are consequences of the spatial variation 
of the source term and are not impacted by the resolution or parcel number. 
For example, the integral of $f$ over the domain is simply the total amount 
of the source, mass, momentum, or energy, transferred per unit time into the
Eulerian phase.

The main focus of our attention is the asymptotic behavior of the expression for small 
$\Delta x$.  In such situations the statistical error represented by the 
first term dominates because the exponent on $\Delta x$, $a-d$, is usually 
less than the exponent 4 that is present in the spatial error term.  
Contrasting Eqn.~\ref{eq:L2} to Eqn.~\ref{eq:mse_nn} provides some interpretation 
of this statistical error term.  We see in Eqn.~\ref{eq:mse_nn} the typical 
$1/\sqrt{n}$ error that is evident in Monte-Carlo sampling calculations.  
However, when we recognize that the sample size must be tied to the mesh 
resolution for the nearest-node kernel, the behavior of this term changes.  In 
Eqn.~\ref{eq:L2} we see that $n$ has been replaced by a dependence on $\Delta 
x$.  The dependence on dimension is also a consequence of the definition of 
source terms -- sources are an intensity: a source per length, per area, or per 
volume.

The rightmost term under the radical in Eqn.~\ref{eq:L2} is not sufficent for 
second-order spatial accuracy.  For CFD codes that do not meet the requirements 
for second-order spatial accuracy, the second order spatial error in Eqn.~\ref{eq:L2} 
may be over-shadowed by other first-order errors in other parts of the spray 
calculation.  For simulations involving very large numbers of parcels or coarse mesh 
resolution, the spatial error will dominate.  The work of Are et al.\ demonstrated the 
second-order convergence of two-dimensional spray calculations in the limit of large 
numbers of parcels \cite{Are}.
 
The convergence of the first term and the second term in Eqn.~\ref{eq:L2} are 
different, with the rate of convergence at very small $\Delta x$ usually 
limited by the first term for reasons of computational cost.  In the well-resolved 
limit, we neglect the second term and observe that the power law behavior of the 
$L_2$ norm as a function of $\Delta x$ is one-half of $a-d$. If we denote the power 
law behavior of $L_2$ as the convergence rate $c$, then the following rule is 
deduced for $a \geq d$.

\begin{equation}
 \label{eq:rule}
 a = 2c + d
\end{equation}

Put simply, for a convergence rate $c$ in a $d$ dimensional calculation of the 
sources \textit{at a single time step}, the number of parcels is governed by the 
exponent $a$. This expression can be used to calculate the behavior of the number 
of parcels per cell, $n_{pc}$, for a convergence rate $c$.  
\begin{equation}
 \label{eq:npc}
 n_{pc} \propto \frac{1}{\Delta x ^ {2c}}
\end{equation}

One interesting consequence of this expression is that, for the simple kernel 
considered here, if the number of parcels per cell is held constant as the 
mesh resolution is increased, the predicted convergence rate is zero, meaning 
convergence will not occur for a single time step.  The situation of a fixed number 
of parcels per cell corresponds to $a=d$, where the statistical error term in 
Eqn.~\ref{eq:L2} does not diminish. 
 
Physically, this result is a consequence of the source terms being a density 
function, calculated per unit area or volume, and depending on the 
dimensionality 
of the problem.

In general, Eqns.~\ref{eq:rule} and \ref{eq:npc} indicate that the number of 
parcels per cell must increase dramatically in order to observe convergence in 
a multi-dimensional CFD calculation.  For second-order convergence (e.g. $c=2$) 
in a three-dimensional calculation, the exponent $a$ produced by 
Eqn.~\ref{eq:rule} would be 7.  Referring back to Eqn.~\ref{eq:model}, this 
means that the number of parcels must increase with the inverse of $\Delta x$ to 
the seventh power.

\section{A Simplified Demonstration of a Single Time step}

For demonstrating convergence, a test case was employed that permits an 
analytical solution for the sources at an instant in time. In this 
test case, particles were stochastically distributed over a unit square 
with a spatially varying number density proportional to the product of
$sin \left(\pi x\right)$ and $sin\left(\pi y\right)$. 
Each particle contributed a uniform amount to the source term calculation, but 
the spatial variation in number density created a source as a function of 
$x$ and $y$ \cite{Schmidt2006}.  The number of parcels was governed by 
Eqn.~\ref{eq:rule} in order to test the theory.

The sources were compared to an exact solution and an $L^2$ error norm was 
calculated using an analytical expression. The number of parcels in the 
domain was set according to Eqn.~\ref{eq:model} with $a$ equaling $2,4,6$.

\begin{figure}[h!]
\centering
\includegraphics[width=0.85\textwidth]{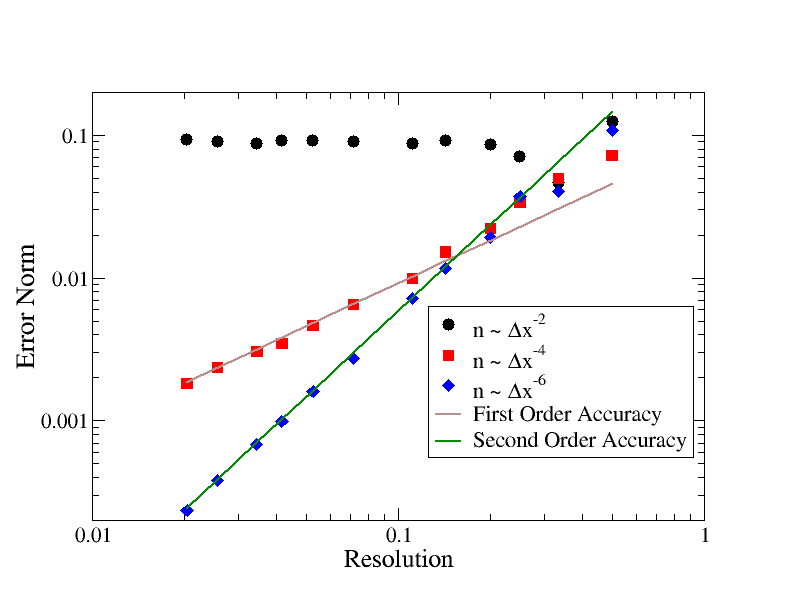}
\caption{Accuracy assessment with a 2D static test.}

\label{fig:2Dtest}
\end{figure}
 
Figure \ref{fig:2Dtest} shows that, as predicted by Eqn.~\ref{eq:rule}, second 
order accuracy requires the number of parcels to increase to the sixth power of 
the reciprocal of mesh spacing.  The first order accuracy with $a=4$ is also 
predicted by Eqn.~\ref{eq:rule}. As previously derived by Schmidt
\cite{Schmidt2006} convergence rates beyond second order are not possible 
unless the kernel weighting changes sign.  For such higher-order kernels, 
however, the statistical errors may become more pronounced. 

Note that for $a=2$, where the number of parcels per cell remains constant, 
the resulting rate of convergence is predicted to be zero, indicating a failure 
to converge.  This prediction is consistent with the empirical results of 
Fig.~\ref{fig:2Dtest}.  Thus it is shown that, unless one uses special 
techniques to moderate statistical error, as in Garg et al.\ \cite{Garg09}, a 
constant number of parcels per cell is not sufficient to obtainin convergence.  
Again, this is consistent with Eqn.~\ref{eq:npc} for a value of $c$ of zero.

The results do not show the characteristic minimum present in Fig.
\ref{fig:optimum}.  The curve plotted in Fig.~\ref{fig:optimum} corresponds to 
the case where $a=0$ and the number of parcels in the entire domain remains 
fixed. Obviously, under such conditions, convergence is not possible.

The other important lesson of this test is the magnitude of the cost of 
achieving convergence.  For the second order convergence test, the number of 
parcels \textit{per cell} increased from eight to over forty-eight million as 
the spatial resolution was increased by a factor of fifty.  As one can see  
from Eqn.~\ref{eq:rule}, the cost of three-dimensional convergence is even 
higher. Fortunately, transient simulations will benefit from the repeated use of 
parcels as they persist over multiple time steps.

\section{Extension to Transient Spray Simulations}

The prior analysis applies to the calculation of sources at a single fixed time. 
There are several transient scenarios that could be considered.  In a 
quasi-steady scenario, the parcels are injected at the same average rate at which
they leave the domain.  Another scenario is of transient injection, where the 
domain initially has no parcels, but rather parcels are injected over a finite time.  

The latter case will be considered here, as it is particularly germane to 
transient spray calculations. It will be assumed that the time of injection is 
less than the parcel residence time, so that no parcels leave the domain. We will 
analyze the cumulative effect of the gas/spray coupling over a time that is 
equal to the time of injection.

The critical extension from the previous analysis to transient sprays is to 
recognize that a parcel is re-used in every subsequent time-step for estimating 
sources.  This produces a sample size that 
is much larger than the number of parcels $n$ injected over the total 
duration of injection.  If a fixed number of parcels $N_{PT}$ is injected 
every time step, after $N_T$ steps the total sample size $s$ becomes the sum 
of a series, as given in Eqn.~\ref{eq:series}.  

\begin{equation}
 \label{eq:series}
 s = N_{PT} + 2N_{PT} + 3N_{PT} ... N_T \cdot N_{PT} 
   \approx \frac{1}{2} N_{PT} N^2_T 
\end{equation} 

This analysis presumes that the quantity of interest is the result of the 
cumulative effect of the transient injection.  Thus the sample size is the 
cumulative sum of parcels in the domain over all time steps.  Hence, the first 
term in the series corresponds to the first time step after the start of 
injection, the second term corresponds to the second time step, etc.

Consider an injection over a duration, $\tau$.  If the Courant-Friedrichs-Lewy 
(CFL) number is constant and equal to unity then the number of time steps 
can be eliminated in the following expression for sample size.  Here, $U$ 
is the velocity scale that appears in the CFL number, which is treated as a 
constant. It is assumed that no parcels exit the calculation, and thus the 
number of parcels $n$ is the product of the number of steps, $N_T$, and $N_{PT}$.

\begin{equation}
 \label{eq:samplesize}
 s = \frac{1}{2} N_{PT} N^2_T 
   = \frac{1}{2}n \cdot N_{T} = \frac{n \tau}{2 \Delta t} = \frac{n \tau U}{2 \Delta x}
\end{equation}

We see that the sample size is no longer simply the number of parcels $n$, but 
is instead a more complicated expression that depends on the mesh resolution.  
This expression is now substituted into Eq. \ref{eq:L2}, where the sample size $s$ 
replaces $n$. The result is more favorable to observing convergence than the single 
time-step result, but still represents considerable expense.
\small
\begin{equation}
 \label{eq:L2transient}
 L_2 \approx \\
 \\
 \sqrt{
 \frac{2 \Delta x^{a-d+1} }{b \tau U} \int_{x \in R^d} f dx^d 
 + 
 \frac{\Delta x^4}{4}  \int_{x \in R^d}  
 \left[
 \sum_{m=1}^{d} \frac{ \partial^2 
f}{\partial x_m^2}
\right]^2
dx^d
 }\\
\end{equation}
\normalsize

As before, for the case where the first term dominates the error, a rule for 
transient CFD convergence can be obtained in the limit of a well-resolved 
simulation. The convergence rate of the $L_2$ norm follows the exponent of the 
first term, one half of the quantity $a-d+1$.  This is rearranged to produce 
Eqn.~\ref{eq:CFDrule}:
\begin{equation}
 \label{eq:CFDrule}
 a = 2c + d -1
\end{equation}

Compared to the results for a single time step, the result of 
Eqn.~\ref{eq:CFDrule} produces a value of $a$ that is reduced by unity.  
Because stochastic fluctuations are averaged over a large number of time 
steps, transient calculations are less expensive to converge.  This result 
assumes the calculation proceeds with a constant time step and invokes 
the Courant number to express the number of steps as a function of $\Delta x$.  
This extra $\Delta x$ factor is responsible for the $-1$ in 
Eqn.~\ref{eq:CFDrule}.  Note also that this improvement in convergence is only 
expected for time-averaged quantities considered for the injection duration $\tau$.

A simple example will elucidate the predictions of this analysis. Consider the 
relatively simple case of refining a three-dimensional mesh containing cubic cells.  
If the cell size is halved, the number of cells increases by eight. To obtain 
first-order convergence in such an oct-tree refinement scheme, 
Eqn.~\ref{eq:CFDrule} indicates that the value of $a$ is four and thus the total 
number of parcels must increase by a factor of sixteen with each level of refinement.
This rule may be expressed more conveniently using the number of parcels per 
cell, $n_{pc}$.  

\begin{equation}
  \label{eq:npc_transient}
 n_{pc} \propto \frac{1}{\Delta x ^ {2c-1}}
\end{equation}

This expression indicates that keeping the number of parcels per cell 
constant as the mesh is refined yields an order one-half convergence rate in 
transient spray simulations.  First order convergence would require a 
doubling of the number of parcels per cell if the cell size is halved.  Second 
order convergence would require increasing the number of parcels per cell by a 
factor of eight.  

Note that a fundamental difference between the single-step result and the 
transient CFD situation is that the strategy of maintaining a constant number 
of parcels per cell will converge for transient calculations.  Repeating the 
thought-experiment for refining cubic cells, if a value of $a=3$ is used, 
Eqn.~\ref{eq:CFDrule} indicates that the rate of convergence will be one-half.  

This result gives a new perspective on the results of Senecal et al.\ 
\cite{Senecal2012}.  In their convergence study, they considered a diesel spray 
emanating from a 90 micron orifice.  They started with a 2 mm mesh resolution 
and applied oct-tree refinement down to a resolution of 31 microns over
seven increasingly refined runs.  For the first five runs, they kept the number 
of parcels per cell constant, thus increasing the number of parcels in the 
computation by a factor of eight.  For the final two runs, they increased the 
number of parcels more slowly, but still used a very large number (21 million 
for the finest mesh).  Their observations indicated that their results were 
converging.

It is likely that the spatial error term in Eqn.~\ref{eq:L2transient} was still 
the limiting term in their computations.  This is a logical consequence of two 
important details in their study.  The first is that even the finest mesh was 
only one-third the initial spray width.  Thus, the spatial error term in 
Eqn.~\ref{eq:L2transient} was significant.  As a comparison, their finest 
resolution would roughly correspond to the coarsest point in 
Fig. \ref{fig:2Dtest}.  Secondly, regardless of the rate of increase, their 
large number of parcels would help keep the size of the first term in 
Eqn.~\ref{eq:L2} relatively small, corresponding to a large value of $b$.  
In any case, the analysis of the present work predicts that their calculations 
should converge.

\section{A Transient Demonstration}

A quantitative test of convergence rate was performed using STAR-CCM+ version 
12.06. An injector was specifically constructed in order to ensure smoothness 
in all spatial distributions.  A typical injection has an abrupt change in 
number 
density at the edge of the orifice where the probability of injecting a parcel 
goes to zero abruptly as the radial location reaches the orifice radius. A probability 
distribution function proportional to the cosine of distance from the injection
center of injection was used to ensure a smooth transition.

For similar reasons, a sine function was used for the mass flow rate curve.  
Additionally the initial time step of each parcel was stochastically varied
immediately after injection in order to uniformly spatially sample the
volume downstream 
of the injection location.  For the sake of simplicity no breakup, turbulence, 
or collision sub-models were used.

The nominal injector radius was 1 $cm$ and the computational domain was a 
2x2x3 $cm$ hexahedron, where the direction of injection was oriented with the 
longest dimension of the domain.  The droplets had a uniform size and a velocity 
of 10 $m/s$, and the duration of injection was 2 $ms$.  The Courant number was 
maintained at a value of unity.  First order time stepping and second order 
spatial discretizations were used.

Two series of convergence tests were run, the first with 100 $\mu m$ diameter 
droplets and the other with 10 $\mu m$ droplets.  Though drop size does not 
appear in the error analysis, altering the drop size changes the integrals that 
appear in Eqn.~\ref{eq:L2transient}.  Smaller droplets correspond to shorter 
deceleration length scales and greater values of velocity second derivatives.

Each series started with 30x30x45 cells and increased the resolution up to 
240x240x360 cells.  The mesh consisted of cubic cells.  Table 1 below shows the 
run matrix for both the larger droplets and smaller droplets.  In each case, 
the value of $b$ was $5.93 \cdot 10^{-10}$ and the expected rate of convergence 
was unity.  The average number of parcels per cell are low--less than unity--due 
to the non-uniformity of the droplet spatial distribution in the domain.  For 
much of the injection duration the majority of the cells in domain do not have 
parcels. Correspondingly, the boundaries were far enough away from the 
injection 
plume that they did not influence the trajectories.

\begin{table}[h]
\begin{center}
\begin{tabular}{|c|r|r|r|r|} \hline

Run & Cell size   & $N_{PT}$ &   n at EOI & n/cells \\
    &  [$\mu m$]  &         &            &         \\ 
\hline
  1 &     667     &   100   & 3,000      & 0.074   \\
  2 &     333     &   800   & 48,000     & 0.148   \\  
  3 &     167     &  6400   & 768,000    & 0.296   \\
  4 &      83     & 51200   & 12,300,000 & 0.593   \\
\hline
\end{tabular}
\end{center}
\caption{Run matrix for both the large and small droplets}
\label{conditions}
\end{table}
 
Two metrics were used to judge convergence.  The first was the penetration of 
the spray center of mass at the end of injection.  This is similar to the 
frequently-noted tip penetration, but less statistically noisy and more clearly 
defined.  The second metric was the gas kinetic energy.  Because the gas was 
initially quiescent, all gas momentum originates with the liquid phase.  Thus, 
the gas kinetic energy is a global measure of the interphase coupling.  
With the parameters of this refinement study, the metrics should vary 
approximately linearly.  Two limitations to this linear convergence are that 
(1) the theory only predicts the mean square error, and so individual 
realizations may vary from the mean and (2) the analysis is only valid when 
$\Delta x$ is sufficiently small compared to all other length scales.

\begin{figure}[]
\captionsetup{margin=10pt,font=small}
\centering
\small
\subfloat[100 $\mu m$ drop size spray center of mass penetration]{
  \includegraphics[width=65mm]{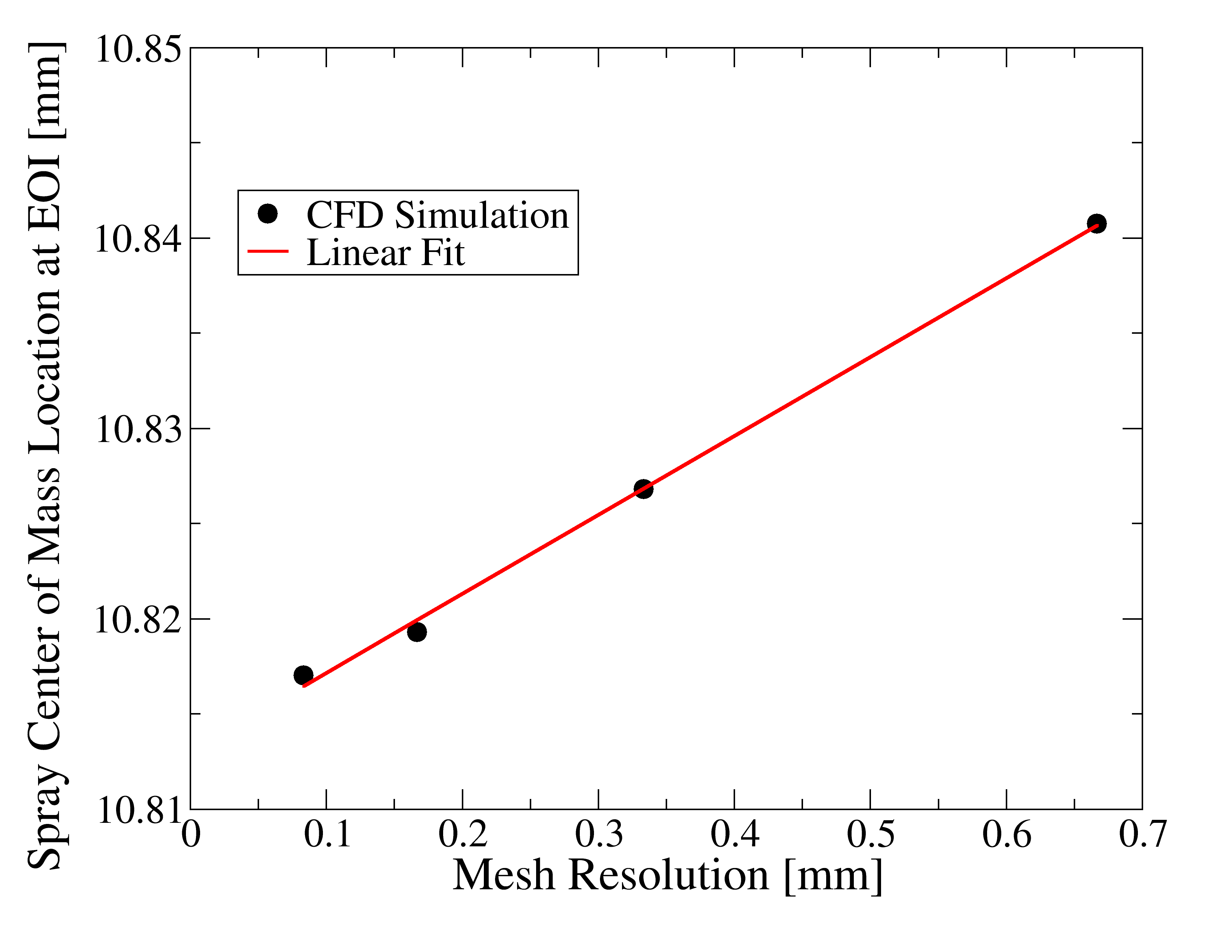}
}
\subfloat[10 $\mu m$ drop size spray center of mass penetration]{
  \includegraphics[width=65mm]{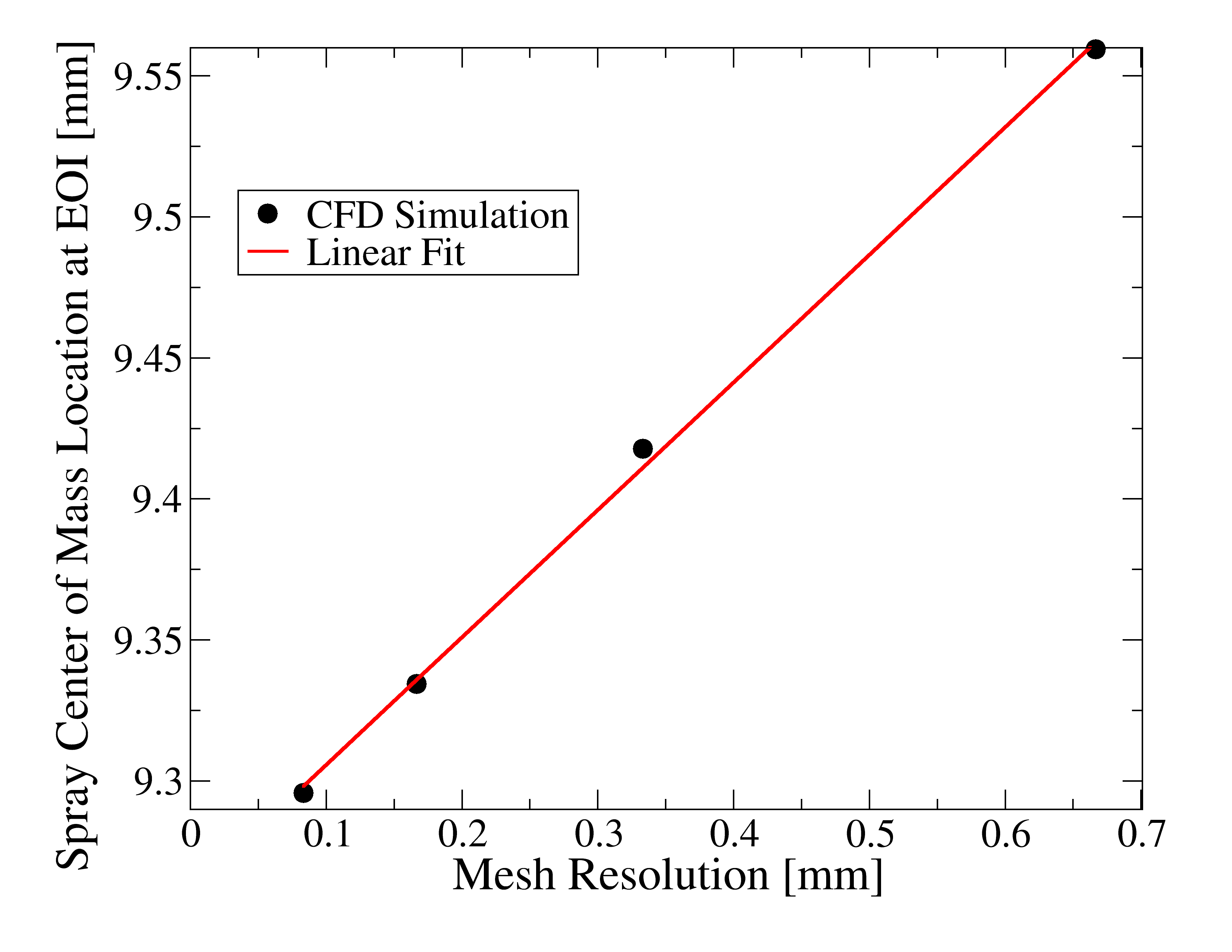}
}
\hspace{2mm}
\subfloat[100 $\mu m$ drop size gas kinetic energy]{
  \includegraphics[width=65mm]{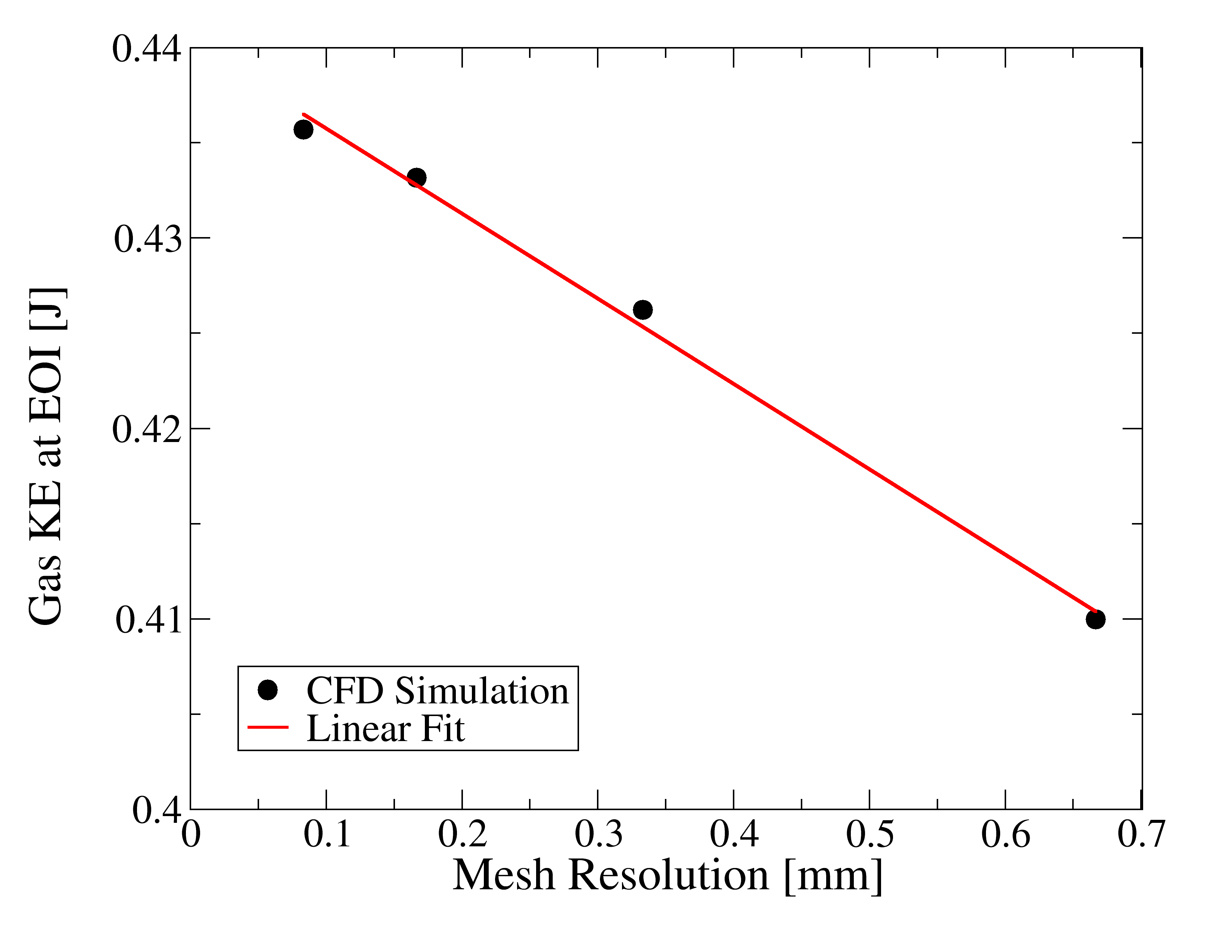}
}
\subfloat[10 $\mu m$ drop size gas kinetic energy]{
  \includegraphics[width=65mm]{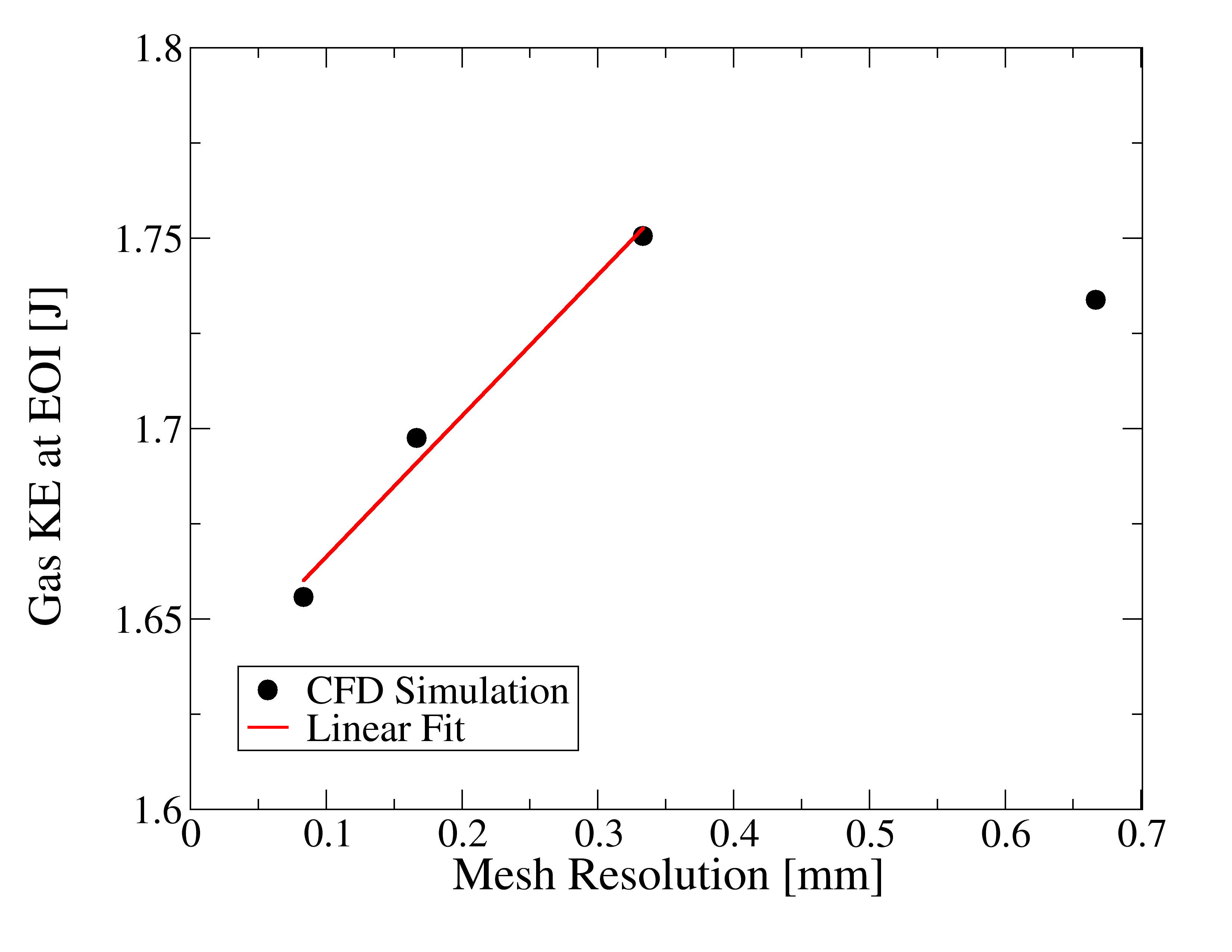}
}
\hspace{2mm}
\caption{Results for uniform drop size convergence tests on spray center of 
mass penetration and gas kinetic energy at the end of injection (EOI). The lines are 
linear regression fits.  In the case of (d), the fit excludes the coarsest mesh 
point.}
\label{convergencePlots}
\end{figure}

These data are presented graphically, as seen in Fig. \ref{convergencePlots}.  
For the conditions simulated here The results shown in plots \ref{convergencePlots} 
(a) through (c) are clearly consistent with with the prediction of linear 
convergence.  The last result in (d) is less clear.  The two anomalous features 
are the non-linear behavior and the change of slope.  The non-linear behavior is 
observed in the coarsest extreme of the curve, which is a consequence of higher 
order error terms being more significant at large $\Delta x$.  The reason that 
convergence is observed for the larger droplets at the coarsest resolution is that 
they have a longer characteristic deceleration length.  The combination of small 
drops and a coarse mesh results in an inadequately resolved simulation. 
The more surprising observation is the different slope between Fig. 
\ref{convergencePlots}(c) and Fig.~\ref{convergencePlots}(d). One would expect 
the slope at the finer points to have the same sign as for the 100 $\mu m$ results.
The differing slope suggests that the physics of the spray behavior has changed 
with different drop sizes.

A third and final test was performed with a more realistic drop size 
distribution.  The injected particle sizes were sampled from a  
Rosin-Rammler distribution.  The cumulative distribution function for the drop 
size $D$ is given by Eqn.~\ref{eq:RRcum}.

\begin{equation}
 f\left(D\right) = 1-exp\left[-\left(D/D_{ref}\right)^\kappa \right] 
 \label{eq:RRcum}
\end{equation}

\begin{figure}[h]
\captionsetup{margin=10pt,font=small}
\centering
\small
\subfloat[Spray center of mass penetration]{
  \includegraphics[width=65mm]{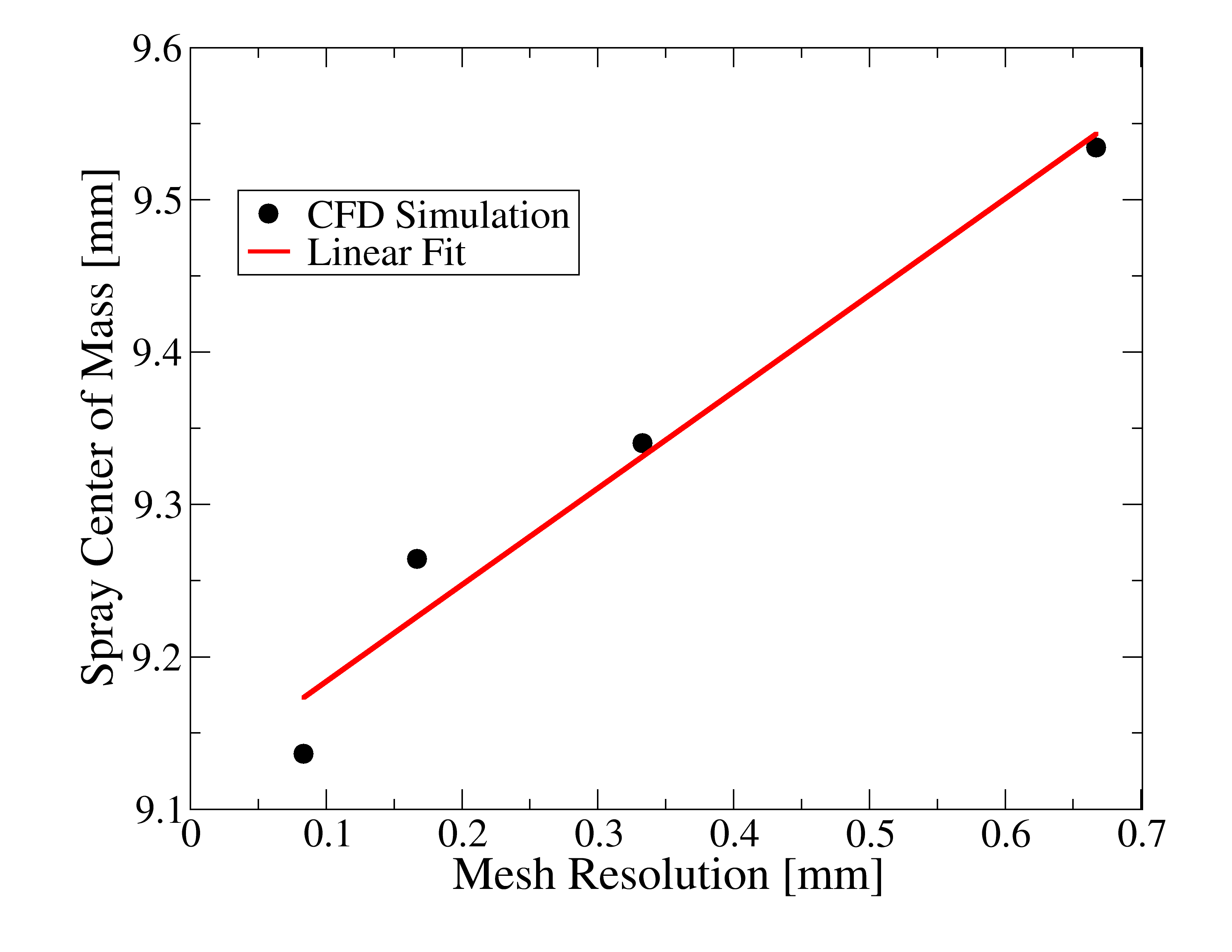}
}
\subfloat[Gas kinetic energy]{
  \includegraphics[width=65mm]{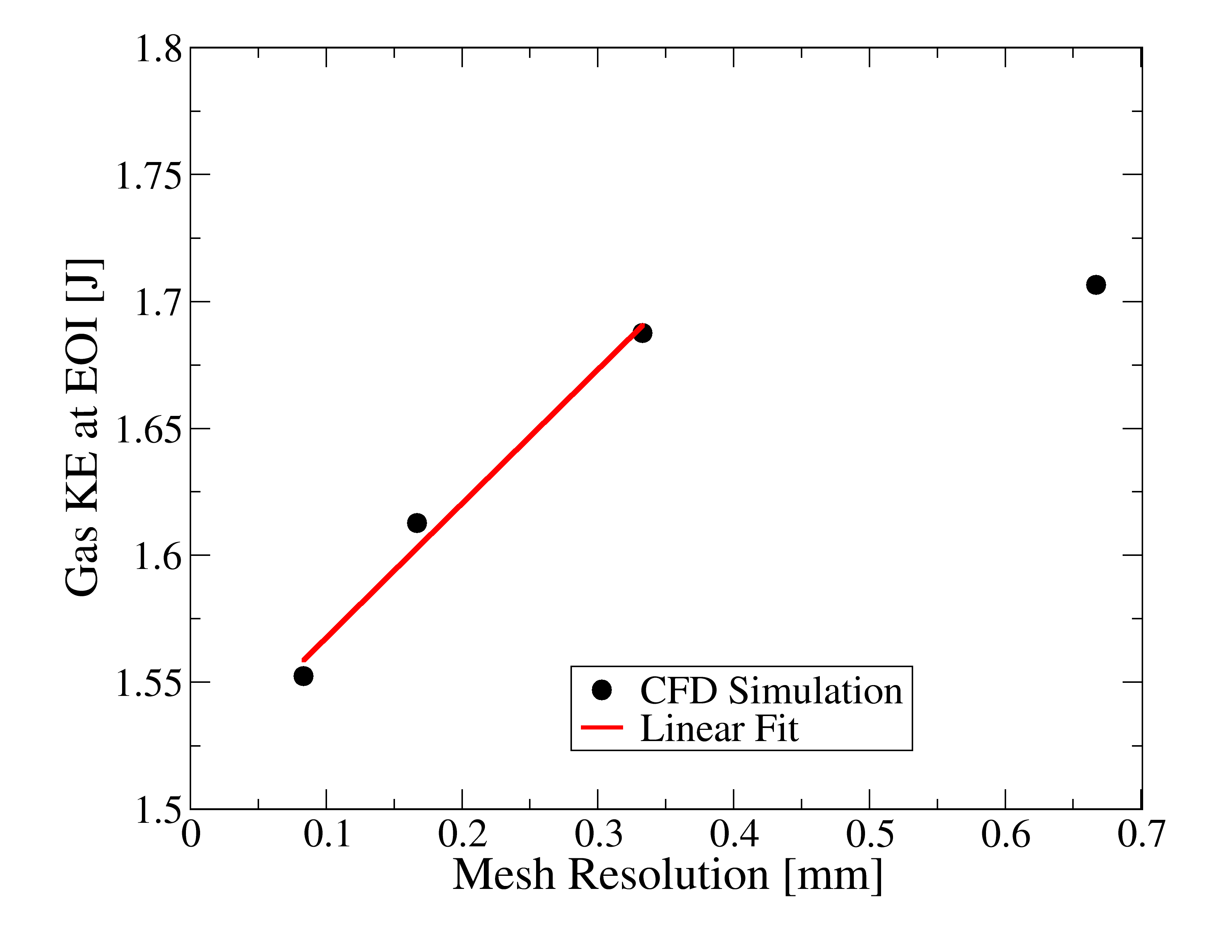}
}

\caption{Results for convergence tests with a Rosin-Rammler drop size 
distribution.  The lines are linear regression fits.  In the case of (b), the 
fit excludes the coarsest mesh point.}
\label{RRconvergencePlots}
\end{figure}

The reference diameter $D_{ref}$ was chosen to be 10 $\mu m$ and the exponent 
$\kappa$ was set to a value of 2.0.  The drop size distribution was limited to 
a range of 1 to 100 microns.  The convergence test was repeated, and the 
results are shown in Fig.~\ref{RRconvergencePlots}.  Not surprisingly, the 
results are similar to the 10 micron mono-disperse case.

Again, the results confirm that the calculations are demonstrating linear 
convergence.  However, the sensitivity inherent in calculating velocity squared 
makes spray center of mass a better metric than gas kinetic energy.  The kinetic 
energy is a square of a 2-norm and is inherently slower to converge than the 
spray center of mass, which is a simple arithmetic mean.

As part of the present work, similar convergence studies were also run with 
FOAM-extend 3.2 and no convergence was observed.  In contrast, Gong 
et al.\ \cite{Gong2010} observed convergence in a modified OpenFOAM solver.
Presumably there are differences between these two code distributions that 
have impact on numerical error.

The forgoing analysis and testing was specific to the case of transient 
injection, where the number of parcels increases at a constant, predictable 
rate.  The analysis applies specifically to a time-averaged quantity that 
originates with the spray and accumulates throughout the development of the spray.
In this case, for example, the gas was initially quiescent, and so the kinetic 
energy is entirely attributable to the cumulative effect of the drag source terms.

\FloatBarrier

\section{Conclusions}
The forgoing paper has defined a simple rule for Lagrangian/Eulerian 
simulations indicating how the number of parcels should vary as the mesh is 
refined in order to observe convergence. The analysis showed that transient 
spray simulations benefit greatly from the fact that parcels are used over 
multiple time-steps.  Empirical tests were consistent with theory presenting 
convergence in three-dimensional tests.  It was noted that a mean center of mass 
spray penetration provided a more reliable metric for judging convergence than  
gas kinetic energy.

The theory was also consistent with prior observations in the literature 
\cite{Senecal2012}. The requirements of high resolution and large numbers of 
parcels helps explain why achieving convergence of Lagrangian/Eulerian 
simulations has been so problematic, resulting in only a handful of successful 
empirical demonstrations \cite{Som2013}.  Presumably, these convergence rules 
can be used as quality assurance tests for spray CFD codes.  When a CFD code 
fails to converge under circumstances where the theory indicates that it should, 
the code should be scrutinized for errors.

To facilitate convergence and reduce the computational cost, future high-fidelity 
Lagrangian/Eulerian simulations may employ some means of ameliorating statistical 
error, similar to Garg et al.\ \cite{Garg09}.  At a minimum, spray simulations 
can take advantage of more sophisticated kernels than the standard approach, 
which is essentially a binary switch. Also, future work could investigate more 
of the sub-models frequently used in spray simulations, such as droplet breakup, 
turbulence interaction, and droplet collision. 

Other areas of future work are the potential importance of other time scales.  
Because stochastic fluctuations in number density give rise to statistical 
error, the time scale over which local number density remains correlated may be 
significant.  For example, a cell may have a positive fluctuation in number 
density that persists for numerous subsequent timesteps.  This duration is a 
potential factor in future analyses.  

The temporal analysis focused on the case of a transient injection, where the 
injection starts at time zero and no parcels left the domain.  As a consequence 
of these assumptions, the transient theory may not be general and may require 
future extension.  Another frequent case is where the injection is quasi-steady 
and the creation of new parcels is balanced, on average, by the departure of 
parcels.  The challenge here is that there is a new timescale that may factor 
into the analysis: the average duration of a parcel within the domain.  
By Little's law, the number of parcels in the domain is directly connected to 
the residence time of the parcel in the domain and the parcel injection rate 
\cite{little2008}.  Because of the prevalence of such quasi-steady simulations, 
this is an intriguing area for future research.

\section*{Acknowledgements}
This work used the Extreme Science
and Engineering Discovery Environment (XSEDE), which is supported by National 
Science Foundation grant number OCI 1053575. The authors acknowledge the Texas 
Advanced Computing Center (TACC) at The University of Texas at Austin
for providing high performance computing resources that have
contributed to the research results reported within this paper.  We also 
acknowledge the use of the Bruce computing resource at Siemens Product 
Lifecycle Management Software, Inc., Lebanon, NH.  We thank Siemens Product 
Lifecycle Management Software, Inc. for their support for this work.

\bibliographystyle{elsarticle-num} 
\bibliography{refs}

\begin{thebibliography}{10}
\expandafter\ifx\csname url\endcsname\relax
  \def\url#1{\texttt{#1}}\fi
\expandafter\ifx\csname urlprefix\endcsname\relax\def\urlprefix{URL }\fi
\expandafter\ifx\csname href\endcsname\relax
  \def\href#1#2{#2} \def\path#1{#1}\fi

\bibitem{Are}
S.~Are, S.~Hou, D.~P. Schmidt, {Second-order spatial accuracy in
  {Lagrangian}--{Eulerian} spray calculations}, Numerical Heat Transfer, Part B
  48~(1) (2005) 25--44.

\bibitem{Schmidt2000}
D.~P. Schmidt, C.~Rutland, A new droplet collision algorithm, Journal of
  Computational Physics 164~(1) (2000) 62--80.

\bibitem{Abani2008}
N.~Abani, A.~Munnannur, R.~D. Reitz, Reduction of numerical parameter
  dependencies in diesel spray models, Journal of Engineering for Gas Turbines
  and Power 130~(3) (2008) 032809.

\bibitem{Subramaniam}
S.~Subramaniam, P.~O'Rourke, {Numerical convergence of the {KIVA}-3 code for
  sprays and its implications for modeling}, {Los Alamos Laboratory Report
  UR-98-5465, Los Alamos, NM}.

\bibitem{Schmidt2006}
D.~P. Schmidt, {Theoretical analysis for achieving high-order spatial accuracy
  in {Lagrangian}/{Eulerian} source terms}, International Journal for Numerical
  Methods in Fluids 52~(8) (2006) 843--865.

\bibitem{stalsberg2004}
K.~Stalsberg-Zarling, K.~Feigl, F.~X. Tanner, M.~Larmi, Momentum coupling by
  means of {Lagrange} polynomials in the {CFD} simulation of high-velocity
  dense sprays, no. 2004-01-0535, SAE World Congress, Detroit, Michigan, 2004.

\bibitem{Garg07}
R.~Garg, C.~Narayanan, D.~Lakehal, S.~Subramaniam, {Accurate numerical
  estimation of interphase momentum transfer in Lagrangian--Eulerian
  simulations of dispersed two-phase flows}, International Journal of
  Multiphase Flow 33~(12) (2007) 1337--1364.

\bibitem{Pope}
T.~Dreeben, S.~Pope, Nonparametric estimation of mean fields with application
  to particle methods for turbulent flows, Tech. Rep. FDA 92-13, {Sibley School
  of Mechanical and Aerospace Engineering, Cornell University}, Ithaca, NY
  14853 (November 1992).

\bibitem{Garg09}
R.~Garg, C.~Narayanan, S.~Subramaniam, {A numerically convergent
  {Lagrangian}--{Eulerian} simulation method for dispersed two-phase flows},
  International Journal of Multiphase Flow 35~(4) (2009) 376--388.

\bibitem{Senecal2012}
P.~Senecal, E.~Pomraning, K.~Richards, S.~Som, Grid-convergent spray models for
  internal combustion engine {CFD} simulations, in: ASME 2012 Internal
  Combustion Engine Division Fall Technical Conference, American Society of
  Mechanical Engineers, 2012, pp. 697--710.

\bibitem{Pai}
M.~G. Pai, S.~Subramaniam, A comprehensive probability density function
  formalism for multiphase flows, Journal of Fluid Mechanics 628~(181) (2009)
  102.

\bibitem{Epanechnikov}
V.~A. Epanechnikov, Non-parametric estimation of a multivariate probability
  density, Theory of Probability \& Its Applications 14~(1) (1969) 153--158.

\bibitem{Rosenblatt}
M.~Rosenblatt, et~al., Remarks on some nonparametric estimates of a density
  function, The Annals of Mathematical Statistics 27~(3) (1956) 832--837.

\bibitem{Gong2010}
Y.~Gong, F.~Tanner, O.~Kaario, M.~Larmi, Large eddy simulations of hydrotreated
  vegetable oil sprays using {OpenFOAM}, in: International multidimensional
  engine modeling meeting at the SAE Congress, Detroit, Michigan, 2010.

\bibitem{Som2013}
S.~Som, D.~Longman, S.~Aithal, R.~Bair, M.~Garc{\'\i}a, S.~Quan, K.~Richards,
  P.~Senecal, T.~Shethaji, M.~Weber, A numerical investigation on scalability
  and grid convergence of internal combustion engine simulations, Tech. rep.,
  SAE Technical Paper (2013).

\bibitem{little2008}
J.~D. Little, S.~C. Graves, Little's law, in: Building intuition, Springer,
  2008, pp. 81--100.

\end{thebibliography}

\end{document}